# Distinguishing Dynamic Phase Catalysis in Cu based nanostructures under Reverse Water Gas Shift Reaction

*Ravi Teja Addanki Tirumala[⊥], Sundaram Bhardwaj Ramakrishnan[⊥], Marimuthu Andiappan*, [⊥]*

**Affiliations:**

[⊥] School of Chemical Engineering, Oklahoma State University, Stillwater, OK, USA.

* **Corresponding Authors,** Email: mari.andiappan@okstate.edu;

ABSTRACT

Increasing anthropogenic carbon dioxide ($CO_2$) emissions have led to rising global temperatures and climate change. Using earth-abundant metal-oxide catalysts such as $Cu_2O$ for reducing $CO_2$ through RWGS reaction seems lucrative. In this work, we have used $Cu_2O$ nanostructures and identified its activity, stability, and selectivity for reducing $CO_2$ to carbon monoxide (CO) which can be further hydrogenated to higher hydrocarbons using Fisher Tropsch synthesis. We have observed that the rate of $CO_2$ conversion increases by 4 times and significantly drops at 300 ºC where the catalyst was reduced to metallic Cu and the rate increases slightly as the temperature is further increased. The selectivity of $CO_2$ reduction is majorly towards CO with a trace amount of methane. We can further exploit the Mie resonance characteristics of $Cu_2O$ nanocatalysts and in-situ generation of hydrogen for hydrogenation of $CO_2$ to enhance the activity of the catalysts. We can further identify the optimum size and shape of the nanocatalysts required and use hybrid nanostructures which can favor RWGS reaction thus improving the stability of these catalysts.

INTRODUCTION

In the last 50 years, concentrations of $CO_2$ emissions have increased exponentially, which has led to consequences like the rise in global temperatures[1]. This rise in temperatures are attributed to climate change which in turn is closely connected to the unsustainable anthropogenic consumption of fossil fuels[2]. Fossil fuels like petroleum, coal, natural gas occupied about 84% of the total world energy mix in 2022[3]. This challenge led to shifting global energy policies[4–6] towards sustainable energy generation through the development of renewable energy solutions.

The conversion of $CO_2$ to CO is considered to be an important approach to the mitigation of $CO_2$ emissions.[7] Commercially, $CO_2$ is converted to CO using the reverse water-gas shift (RWGS) reaction. The formation of CO from this reaction is the building block, to be further hydrogenated to higher hydrocarbons which can be used for energy generation and applications through the Fisher–Tropsch synthesis, and methanol synthesis[7–9]. RWGS reaction can be used in the production of ethylene glycol, styrene, and light olefins[7,9]. Though direct methanation of $CO_2$ is thermodynamically favorable, hydrogenation of CO can improve the yield of methane and adds to the versatility of CO transformations to various high-value hydrocarbons and fine chemicals.[9] using renewable generated hydrogen and photocatalytic[10], electrochemical approaches can reduce the footprint of the reaction further.[7,9] Thermodynamically RWGS reaction is favored at higher temperatures and it increases the hydrogen required for the reaction by an extra mole, in contrary to direct methanation of $CO_2$.[7–9,11] Various supported metal and bimetallic catalysts such as Cu,Pd, Pt, Rh, Ni supported on $CeO_2$,$SiO_2$, $Al_2O_3$ were investigated.[9,11,12] Through chemical looping of RWGS reaction, metal oxide catalyst is first reduced by hydrogen and the active catalyst is oxidized from $CO_2$ with no excess hydrogen required, making the process more efficient.[7–9]

In this contribution, we have identified the dynamic phase changes occurring on Cu-based catalysts across the RWGS reaction and performed these reactions with different shapes of $Cu_2O$ nanostructures. $Cu_2O$ nanoparticles were synthesized using microemulsion technique where n-heptane is the oil phase. Copper nitrate ($Cu(NO_3)_2$) is used as a precursor and reduced with hydrazine solution. Brij L4 is used as a surfactant which allows to form uniform sized nanoreactors where Cu nanoparticles grow with time. Acetone is used to break this emulsion and further these nanoparticles are washed multiple times to obtain $Cu_2O$ spheres of size $34 \pm 4$ nm.

Chemical Reduction method is used to synthesize Cubic $Cu_2O$ nanostructures where $CuCl_2$ is used as the precursor and reduced with NaOH and Sodium Ascorbate, in water as a medium. This mixture is allowed to stir and 600 rpm for an hour and these nanoparticles formed are washed in ethanol. The size of these nanoparticles with an edge length of 325 ± 37 nm, were obtained using SEM.

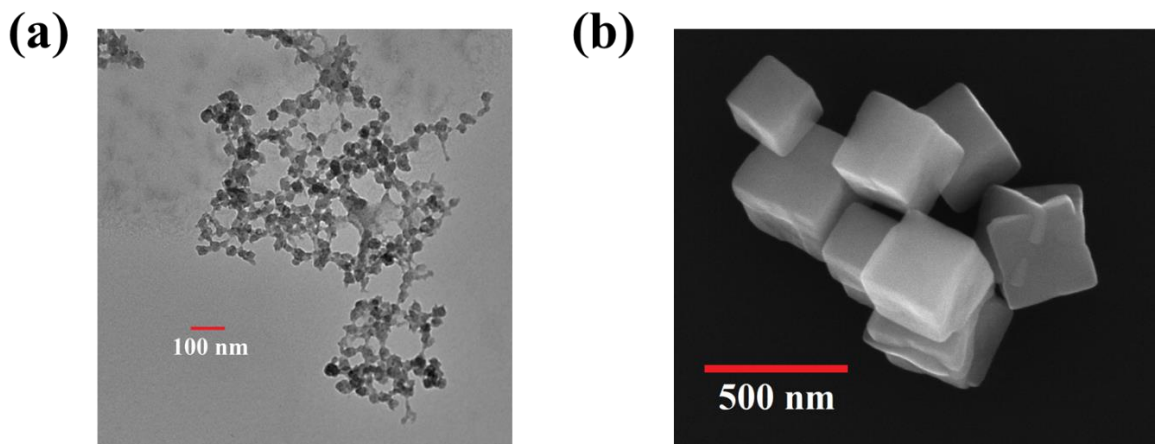

**Figure 1** (a) Representative TEM image of Spherical $Cu_2O$ nanoparticles with diameter of 34 ± 4 nm (b) Representative SEM image of Cubic $Cu_2O$ nanoparticles with an edge length of 325 ± 37 nm

$Cu_2O$ spheres with diameter of 34 ± 4 nm is supported on Silica (2wt%) and 20 mg of which was loaded into the high temperature reaction chamber (Harrick), and a total flow of 100 mL/min consisting of 95% $CO_2$ and 5% $H_2$ were flown. Initially the gas was allowed to pass through the catalyst bed for 30 minutes to purge the system with any trace atmospheric gases. This reaction was performed up to 500 ºC and the rate of reaction was observed using Shimadzu GC-MS. As observed in Figure 2, the rate of reaction reaches a maxima at 300 ºC and then drops drastically suggesting phase changes on the catalyst. It was also observed using GC-MS the product consists majorly of CO, being the product of interest and trace amount of $CH_4$.

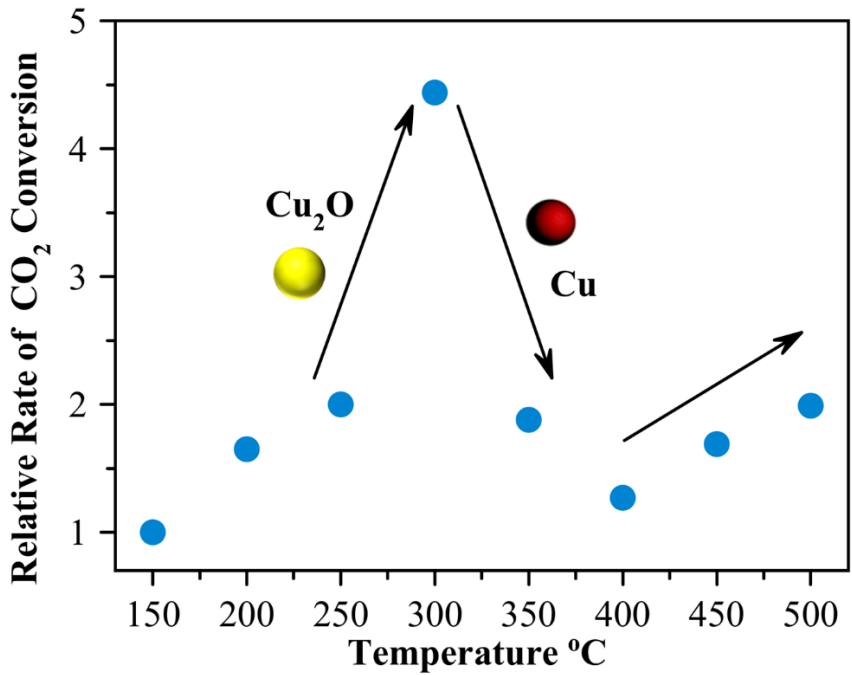

**Figure 2** Relative rate of RWGS reaction as a function of temperature using $Cu_2O$ nanospheres particle diameter of 34 ± 4 nm, with rate identified at 150 °C as reference.

X-ray diffraction patterns suggests the change in the phases of the catalysts before and after reaction as shown in Figure 3. It is also observed that the color of the catalyst changes from a greenish hue to a reddish in color suggesting hydrogenation of $Cu_2O$. This gives us insights into the drop-in activity of the catalyst at 300 °C. This suggests us that $Cu_2O$ supported on silica was completely reduced to Cu by 300 °C and the activity dropped as the metallic Cu was not performing high activity in these thermal conditions as shown in Figure 2. It is observed that rate of reaction slightly raises across 350 °C to 500 °C, attributed to increase in the thermal energy.

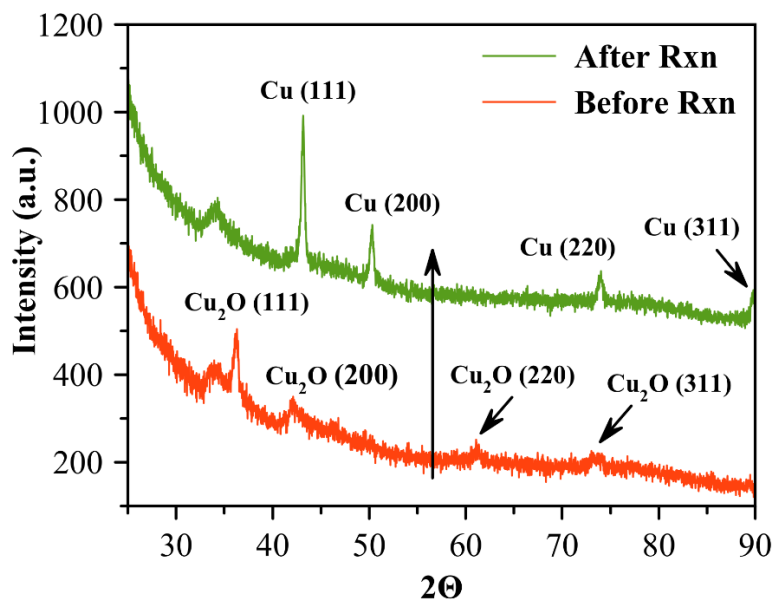

**Figure 3** Representative X-ray diffraction patterns of as-prepared $Cu_2O$ nanospheres before reaction and after exposure to the conditions of the reaction.

Operando diffused reflectance UV-Vis extinction measurements were made using the praying mantis set up (Thermo Scientific) and dynamic changes occurring across the reaction were observed. UV-Vis extinction measurements above 300 ºC as significant increase in the noise in the spectra. This is attributed to significant increase in the vibrations at the atomic level. The reactor was allowed to cool down to room temperature with the same composition of flow rate and an after-reaction UV-Vis was measured suggesting $Cu_2O$ (≈ 490 nm) has been completely reduced to Cu (≈ 580 nm). The dynamic phase change mechanism that $Cu_2O$ (111) facet of nanospheres is reduced to Cu (111) across the reaction, is unidirectional in nature, where Cu(111) nucleates[13], and complete reduction of $Cu_2O$ occurs at 300 ºC.

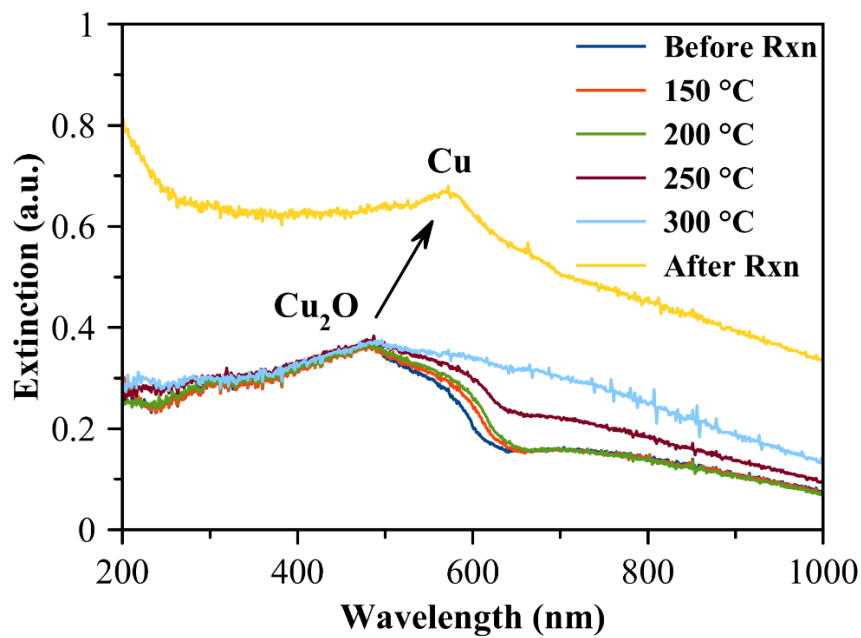

**Figure 4** Operando diffused reflectance UV-Vis extinction spectra of cubic $Cu_2O$ nanoparticles supported on Silica ($SiO_2$)

CONCLUSIONS

Dynamic changes occurring on $Cu_2O$ nanostructures were observed for RWGS reaction and found the conditions at which these phase changes occur has been identified. Using SEM and TEM size and shape of these particles are determined, and phase changes of these particles were observed using XRD technique. Multiple in-line characterization techniques such as GC-MS, praying mantis set-up for diffused reflectance spectra measurements to gain insights into the stability, activity of these catalysts were observed and found that $Cu_2O$ was completely reduced to Cu under the reaction conditions.

We can further exploit the optical characteristics of both $Cu_2O$ to enhance the activity of these catalysts and perform RWGS reaction at lower temperatures mitigating the effects of hydrogenation of the catalyst itself for improving the stability of the catalysts. In-situ generation of hydrogen for use in RWGS reaction can also be explored to mitigate the metal oxide catalysts being reduced.

ACKNOWLEDGEMENT

Dr. Andiappan gratefully acknowledges the funding support from the National Science Foundation (NSF)-CBET CATALYSIS program under Grant No. 2102238. The research results obtained and discussed in this publication were made possible in part by funding through the award for project number HR18-093, from the Oklahoma Center for the Advancement of Science and Technology. The TEM and SEM images were acquired at the Oklahoma State University (OSU) Microscopy Laboratory. We would like to thank Dr. Jim Puckette for the permission to use the XRD facility at Nobel Research Centre, Oklahoma State University.

REFERENCES

(1) Meinshausen, M.; Meinshausen, N.; Hare, W.; Raper, S. C. B.; Frieler, K.; Knutti, R.; Frame, D. J.; Allen, M. R. Greenhouse-Gas Emission Targets for Limiting Global Warming to 2 °C. *Nature* **2009**, *458* (7242), 1158–1162. https://doi.org/10.1038/nature08017.
(2) *Climate Change 2014: Synthesis Report*; Pachauri, R. K., Mayer, L., Intergovernmental Panel on Climate Change, Eds.; Intergovernmental Panel on Climate Change: Geneva, Switzerland, 2015.
(3) Statistical Review of World Energy 2022, 71st Edition. Bp 2022.
(4) Nel, W. P.; Cooper, C. J. Implications of Fossil Fuel Constraints on Economic Growth and Global Warming. *Energy Policy* **2009**, *37* (1), 166–180. https://doi.org/10.1016/j.enpol.2008.08.013.
(5) Palacios, A.; Barreneche, C.; Navarro, M. E.; Ding, Y. Thermal Energy Storage Technologies for Concentrated Solar Power – A Review from a Materials Perspective. *Renewable Energy* **2020**, *156*, 1244–1265. https://doi.org/10.1016/j.renene.2019.10.127.
(6) Basit, M. A.; Dilshad, S.; Badar, R.; Rehman, S. M. S. ur. Limitations, Challenges, and Solution Approaches in Grid-Connected Renewable Energy Systems. *International Journal of Energy Research* **2020**, *44* (6), 4132–4162. https://doi.org/10.1002/er.5033.
(7) Wang, W.; Wang, S.; Ma, X.; Gong, J. Recent Advances in Catalytic Hydrogenation of Carbon Dioxide. *Chem. Soc. Rev.* **2011**, *40* (7), 3703–3727. https://doi.org/10.1039/C1CS15008A.
(8) Gao, S.; Lin, Y.; Jiao, X.; Sun, Y.; Luo, Q.; Zhang, W.; Li, D.; Yang, J.; Xie, Y. Partially Oxidized Atomic Cobalt Layers for Carbon Dioxide Electroreduction to Liquid Fuel. *Nature* **2016**, *529* (7584), 68–71. https://doi.org/10.1038/nature16455.
(9) Daza, Y. A.; Kuhn, J. N. $CO_2$ Conversion by Reverse Water Gas Shift Catalysis: Comparison of Catalysts, Mechanisms and Their Consequences for $CO_2$ Conversion to Liquid Fuels. *RSC Adv.* **2016**, *6* (55), 49675–49691. https://doi.org/10.1039/C6RA05414E.

(10) Wan, L.; Zhou, Q.; Wang, X.; Wood, T. E.; Wang, L.; Duchesne, P. N.; Guo, J.; Yan, X.; Xia, M.; Li, Y. F.; Jelle, A. A.; Ulmer, U.; Jia, J.; Li, T.; Sun, W.; Ozin, G. A. Cu2O Nanocubes with Mixed Oxidation-State Facets for (Photo)Catalytic Hydrogenation of Carbon Dioxide. *Nat Catal* **2019**, *2* (10), 889–898. https://doi.org/10.1038/s41929-019-0338-z.

(11) Zakharova, A.; Iqbal, M. W.; Madadian, E.; Simakov, D. S. A. Reverse Microemulsion-Synthesized High-Surface-Area Cu/γ-Al$_2$O$_3$ Catalyst for CO$_2$ Conversion via Reverse Water Gas Shift. *ACS Appl. Mater. Interfaces* **2022**, *14* (19), 22082–22094. https://doi.org/10.1021/acsami.2c01959.

(12) Kim, J. Y.; Rodriguez, J. A.; Hanson, J. C.; Frenkel, A. I.; Lee, P. L. Reduction of CuO and Cu$_2$O with H$_2$: H Embedding and Kinetic Effects in the Formation of Suboxides. *J. Am. Chem. Soc.* **2003**, *125* (35), 10684–10692. https://doi.org/10.1021/ja0301673.

(13) LaGrow, A. P.; Ward, M. R.; Lloyd, D. C.; Gai, P. L.; Boyes, E. D. Visualizing the Cu/Cu$_2$O Interface Transition in Nanoparticles with Environmental Scanning Transmission Electron Microscopy. *J. Am. Chem. Soc.* **2017**, *139* (1), 179–185. https://doi.org/10.1021/jacs.6b08842.